# Gate-Tunable Mid-Infrared Electroluminescence from Te/MoS$_2$ p–n Heterojunctions

Shiyu Wang[1], Delang Liang[1,2], Zhi Zheng[3], Mingyang Qin[1], Yuchun Chen[1], Jie Sheng[1], Shula Chen[2], Lin Li[3], Changgan Zeng[3], Anlian Pan[2,7†], Jinluo Cheng[4,8†], and Dong Sun[1,5,6,8†]

[1]International Center for Quantum Materials, School of Physics, Peking University, Beijing, China.
[2]Key Laboratory for Micro-Nano Physics and Technology of Hunan Province, Hunan Institute of Optoelectronic Integration, College of Materials Science and Engineering, Hunan University, Changsha, China.
[3]CAS Key Laboratory of Strongly Coupled Quantum Matter Physics, and Department of Physics, University of Science and Technology of China, Hefei, China.
[4]GPL Photonics Laboratory, State Key Laboratory of Luminescence Science and Applications, Changchun Institute of Optics Fine Mechanics and Physics, Chinese Academy of Sciences, Changchun, China.
[5]Collaborative Innovation Center of Quantum Matter, Beijing, China.
[6]Frontiers Science Center for Nano-optoelectronics, School of Physics, Peking University, Beijing, China.
[7]School of Physics and Electronics, Hunan Normal University, Changsha, China.
[8]School of Physics and Laboratory of Zhongyuan Light, Zhengzhou University, Zhengzhou, China.

[†]Email: sundong@pku.edu.cn; jlcheng@ciomp.ac.cn; anlian.pan@hnu.edu.cn;

**Abstract**

Mid-infrared (MIR) emitters are critical components in advanced photonic systems, driving progress in fields such as chemical sensing, environmental monitoring, medical diagnostics, thermal imaging and free-space communications. Conventional MIR emitters based on III-V heterostructures rely on complex epitaxial growth on rigid lattice-matched substrates and suffer from limited integration compatibility with CMOS or flexible platforms. The recent development of novel MIR emitters based on two-dimensional (2D) materials such as black phosphorus (BP) is more suitable for on-chip applications but faces challenges related to stability and emission efficiency. Based on the recently discovered highly efficient photoluminescence of Te, we demonstrate a gate-tunable mid-infrared light-emitting diode based on a van der Waals heterojunction formed by multilayer transition metal dichalcogenide (TMD) $MoS_2$ and tellurium (Te). The device emits polarized electroluminescence (EL) centered at 3.5 μm under forward bias at 25 K, and the EL persists up to 80 K with reduced intensity. Gate control of the $MoS_2$ Fermi level modulates the band alignment and injection efficiency, enabling dynamic tuning of the EL intensity. The emission remains spectrally stable under varying bias and gating, indicating robust band-edge recombination. These results establish the Te/TMD heterostructure as a promising platform for integrated polarized mid-infrared optoelectronics.

**Introduction**

The mid-infrared (MIR, 2.5–25μm) spectral region is essential for numerous applications, including gas sensing, infrared spectroscopy, thermal imaging, and free-space optical communication. Conventional MIR light-emitting technologies have been developed primarily on epitaxial III–V platforms, including InSb/AlInSb quantum-well light-emitting diodes[1], GaInAsSbP alloy light-emitting diodes[2], and intersubband devices based on III–V compounds[3]. Within the same epitaxial technology landscape, type-II InAs/GaSb superlattice interband LEDs established cascaded active-region concepts for 3–4 μm electroluminescence[4] and were subsequently advanced toward higher radiance and output power through optimized cascaded designs, including work from John P. Prineas and co-workers[5]. Interband cascade LEDs further pushed continuous-wave mid-IR LED power into the multi-milliwatt regime, with notable progress from industrial development efforts such as nanoplus[6]. More recently, mercury-chalcogenide colloidal quantum-dot LEDs, including HgTe and HgSe-based systems from Philippe Guyot-Sionnest's group, have emerged as a solution-processable route to mid-IR electroluminescence[7,8]. Despite their strong performance, these established approaches typically rely on epitaxial growth on lattice-matched substrates and device stacks that are less compatible with CMOS back-end processing or mechanically flexible platforms. Moreover, polarization-defined emission and spectral engineering in many mid-IR emitters are commonly implemented by incorporating additional photonic structures such as resonant cavities or gratings, while the growing demand for on-chip mid-IR polarimetry motivates compact sources with intrinsic and electrically controllable polarization. These constraints have spurred interest in alternative optoelectronic systems that combine mechanical flexibility, process scalability, and spectral tunability. Two-dimensional (2D) van der Waals (vdW) semiconductors offer distinct advantages for light emitters[9-15], including atomic thinness, electrostatic tunability, and seamless integration on arbitrary substrates without lattice-matching constraints. Following the isolation of monolayer transition metal dichalcogenides (TMDs), electroluminescence (EL) was rapidly demonstrated in monolayer $MoS_2$[10] and WSe$_2$, whose direct bandgaps and strong excitonic effects enable bright emission in the visible and near-infrared (NIR) regimes. Various device architectures—such as lateral p–n junctions, vertical heterostructures[11], and tunneling diodes[12-14]—have since been developed to realize valley polarization-tunable and gate-controlled EL[15] based on TMDs. Despite these advances, the relatively large bandgaps of TMDs (>1 eV) fundamentally limit their operation wavelength range to the visible–NIR region, thereby limiting their applicability in the mid-infrared region. In addition, light emission from TMDs is not polarized due to the nearly isotropic in-plane optical response of TMDs, and it typically requires external photonic structures to achieve strong linear polarization[16]. This is in contrast to intrinsically anisotropic 2D semiconductors such as black phosphorus or Te. As a result, TMD-based emitters are not ideal for applications in the mid-infrared region.

To overcome these challenges, there has been growing interest in anisotropic 2D semiconductors with narrower bandgaps and polarization-selective optical properties. Black phosphorus (BP) has emerged as a promising MIR emitter, with a thickness-dependent direct bandgap ranging from ~0.3 eV (bulk) to ~2.0 eV (monolayer)[17-19] and pronounced in-plane anisotropy that enables linearly polarized emission[17,20-22]. BP-based MIR light-emitting diodes have been investigated across diverse architectures, including heterojunction emitters, waveguide-integrated implementations, resonant-cavity designs, and van der Waals heterostructure, with recent progresses toward scalable device formats[21-25]. A major impediment, however, is BP's limited ambient stability[26,27], which necessitates effective encapsulation for reliable operation. Encouragingly, long operating lifetimes have been demonstrated for properly encapsulated BP LEDs, highlighting the practical importance of stringent environmental isolation for BP-based MIR emitters. Moreover, BP exhibits strong electric-field-induced spectral tunability via the Stark effect- an attractive route for active wavelength control, yet one that can also introduce emission-energy shifts under vertical electric fields[28]. Recently, tellurium (Te)[29-32], a layered narrow-bandgap (~0.35 eV) p-type semiconductor, has attracted growing attention as an alternative to BP for MIR optoelectronics. Te possesses a thickness-independent bandgap that is nominally indirect with a small k-offset but effectively quasi-direct for optical transitions, along with excellent air stability, high mobility[33,34], and intrinsic linear dichroism[35]. Prior studies have demonstrated MIR photodetection[35-37] and highly efficient photoluminescence[38] in Te-based heterostructures, highlighting its potential for MIR optoelectronics. Very recently, during the preparation of this work, electrically driven mid-infrared electroluminescence has been demonstrated in Te-based van der Waals LEDs that rely on dual-electrode bipolar injection, and the polarization of the emission can be tuned by bias.[39] These advances also link Te-based MIR emitters to Weyl physics, opening research avenues beyond the scope of conventional band-edge optoelectronics. In this work, we develop a gate-tunable heterojunction of Te that enables systematic modulation of the built-in junction electrostatics and interfacial band alignment to control injection and recombination. Based on a vdW p–n heterojunction formed by n-type multilayer MoS$_2$ and p-type Te, as schematically illustrated in Fig. 1a, we demonstrate a mid-infrared light-emitting diode. The band diagram highlights carrier injection and radiative recombination processes localized within Te, giving rise to EL emission centered at 3.5 μm. Gate-tunable carrier modulation enables control over the band alignment and interfacial injection efficiency. The device exhibits linearly polarized EL under forward bias, with the lowest turn-on voltage of ~1.7 V. Spectral analyses confirm that EL arises from direct band-edge transitions in Te, consistent with its PL response. Importantly, the Te/MoS$_2$ platform demonstrates wavelength-stable and polarization-locked MIR emission under electrostatic control, and we further verify its operational robustness through extended storage (~10 months) without the need for strict hermetic encapsulation. These findings establish the Te/MoS$_2$ heterostructure as a promising platform for gate-tunable, polarized mid-infrared light sources in reconfigurable photonic and optoelectronic circuits.

## Results

**Characterization of transport and optical properties**

The device was fabricated via standard mechanical exfoliation and van der Waals (vdW) assembly processes, where multilayer MoS$_2$ was positioned on top of a Te nanosheet. The source and drain electrodes were patterned by electron-beam evaporation of Pd/Au (20 nm/100 nm) and Cr/Au (20 nm/100 nm) onto the Te flake and MoS$_2$ layers, respectively. The 285 nm SiO$_2$ layer serves as the back-gate dielectric, with the p-type silicon substrate used as the back gate. Figure 1b shows an optical micrograph of a typical Te/MoS$_2$ heterostructure, where the boundaries of the MoS$_2$ and Te nanosheets are outlined by dashed blue and green lines, respectively. The Au electrodes are visible as bright metallic regions, outlined by orange dashed lines. To determine the thickness and surface morphology of the MoS$_2$/Te heterojunction, atomic force microscopy (AFM) measurements were performed over the sample area. The thicknesses of MoS$_2$ and Te were measured to be approximately 17 nm and 234 nm, respectively (see Supplementary Information Fig. S1). Raman spectroscopy was performed to confirm the composition and crystalline quality of the heterostructure. For Raman measurement, a 532 nm laser was utilized to excite the characteristic resonant peaks in Te, MoS$_2$, and the heterojunction region. As shown in Fig. 1c, the Te-related peaks at 92, 120, and 140 cm$^{-1}$ are assigned to the E$_1$ (bond-bending mode), A (chain expansion mode), and E$_2$ (bond-stretching mode) phonon vibrational modes; the MoS$_2$-related peaks at 385 and 409 cm$^{-1}$ are assigned to the E$_{2g}$ (in-plane Mo-S vibration mode) and A$_{1g}$ (out-of-plane mode), respectively. In the heterostructure region, all peaks retain their spectral positions without broadening or shifting, demonstrating negligible interlayer strain and chemically intact interfaces. Notably, the absence of additional modes or intensity suppression suggests minimal charge-transfer-induced lattice distortion—a hallmark of high-quality vdW heterojunctions.

We focused on characterizing the device's EL emission under source-drain bias first. To gain deeper insight into the origin of this EL emission, we also carried out photoluminescence (PL) measurements of the Te/MoS$_2$ heterostructure under 1064 nm laser excitation. As shown in Fig. 1e, the resulting PL and EL spectra are presented together for direct comparison. Under forward-bias conditions ($V_g$ = 20 V, $V_{ds}$ = 4 V), the device exhibits EL emission centered at ~3.5 µm. All measurements were performed at 25 K, unless otherwise noted, to enable high signal-to-noise detection of gate-modulated mid-infrared emission. As shown in Supplementary Note 4, electroluminescence persists up to 80 K, although the emission intensity decreases with increasing temperature. A direct spectral comparison with PL reveals that both EL and PL exhibit nearly identical peak positions and spectral linewidths (Fig. 1e), suggesting that both emissions originate from interband transitions of Te, a quasi-direct transition between the valence band maximum and conduction band minimum of Te. Spatially resolved EL mapping (Fig. 1d) shows that the mid-infrared emission primarily originates from the Te/MoS$_2$ overlap (junction) region. The device geometry in Fig. 1d was determined by comparing the EL map to a reflection micrograph of the same area, which is provided in Supplementary Fig. S5.

Given the in-plane anisotropic chain-like crystal structure of Te, the EL emission from Te-based heterostructures is expected to be strongly linearly polarized, similar to the polarized photoluminescence (PL) reported in our recent work [40]. The polarized emission is crucial for polarization-sensitive optoelectronic applications. Fig. 1f presents measurements of the polarization characteristics of the EL emission. Notably, both EL and PL emissions reach maximum intensity when the polarization is aligned along the armchair direction (a-axis) and minimum intensity when the polarization is aligned along the chain direction (c-axis), reflecting the intrinsic optical anisotropy dictated by selection rules of the dipole matrix elements near the Te bandgap, where radiative transitions are preferentially oriented along the a-axis. To quantify the polarization, we use the degree of linear polarization (DOP), defined as $\mathrm{DOP} = (I_{\max} - I_{\min})/(I_{\max} + I_{\min})$. The DOP of the measured device is also bias dependent, consistent with a recent report on graphene-Te-graphene mid-infrared LEDs[39]. The EL polarization plot shown in Fig. 1f yields a DOP of 0.70. While a comprehensive bias- and gate-dependent study of the polarization behavior is beyond the scope of the present work and will be addressed in a separate publication, we note that the polarization state varies significantly across different devices. Different Te flakes in Te/TMD heterojunctions exhibit different degrees of linear polarization in EL, while in an independent Te/WSe$_2$ device, we observe a nearly unity EL DOP (see Supplementary Fig. S6). The possibilities of these complex DOP results may arise from the flake-to-flake variations in the intrinsic doping of Te, which shifts the equilibrium Fermi level and thereby influences the polarization; and recent Te-based MIR LEDs have shown that the DOP can vary with the injection condition[39]. A comprehensive bias- and gate-dependent study of the polarization behavior is beyond the scope of this work and require further investigation. Having established that the Te/MoS$_2$ heterostructure functions as a polarized mid-infrared light emitter, we next examine its electronic transport properties and interfacial band alignment. In the present device platform, Te flakes were selected from hydrothermal batches to ensure sufficient morphological uniformity for reliable junction formation, while few-layer MoS$_2$ was employed to provide robust electron transport and stable current drive for reproducible EL turn-on. A brief rationale for the thickness selection is provided in Supplementary Note 1.

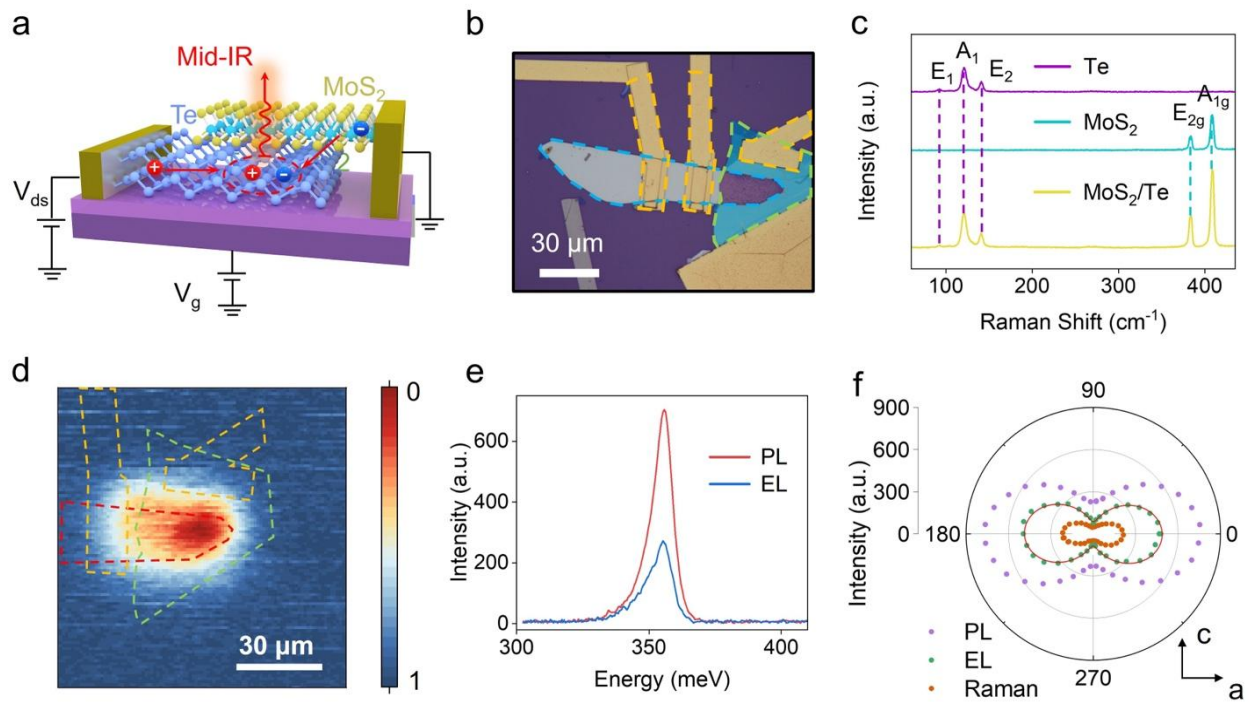

**Fig. 1 | Schematic and optical properties of the Te/MoS$_2$ heterojunction. a,** Schematic illustration of the Te/MoS$_2$ heterojunction device. **b,** Optical micrograph showing the Te/MoS$_2$ heterojunction region (outlined by blue/green dashed lines) on a Si/SiO$_2$ (285 nm) substrate used as the back-gate. Scale bar: 30 μm. **c,** Raman spectra of isolated Te (purple), MoS$_2$ (blue), and the heterojunction region (yellow). **d,** Spatially resolved EL intensity map of the Te/MoS$_2$ heterojunction LED, recorded in the 3.5 μm band under forward bias ($V_\mathrm{g} = 20\,\mathrm{V}, V_\mathrm{ds} = 4\,\mathrm{V}$; 25 K). The Te flake and MoS$_2$ region are outlined by red and green dashed lines, respectively. **e,** PL (red) and EL (blue) spectra of the Te/MoS$_2$ diode. **f,** Polarization characterization of PL, EL and the A$_1$ Raman mode. All curves are fit by $I = (I_{\max} - I_{\min})\cos^2\theta + I_{\min}$, where θ is measured from the

a-axis of Te, and $I_{max}$ and $I_{min}$ are the EL intensities along the a-axis and c-axis, respectively.

The electrical properties of the Te/MoS$_2$ heterostructure were systematically investigated to elucidate its p–n junction behavior and gate-dependent transport characteristics at 25 K. The output characteristics ($I_{ds}$–$V_{ds}$) measured at back-gate voltages ($V_g$) of 30 V, 0 V, and –30 V show pronounced, gate-tunable rectification behavior, as depicted in Fig.3. At $V_g$ = +30 V, when the MoS$_2$ channel is strongly n-doped, the device exhibits clear rectification behavior. At $V_g$ = 0 V, the rectification is strongly suppressed, and the currents remain in the submicroampere range over the entire $V_{ds}$ window. At $V_g$ = –30 V, the current stays at the measurement noise floor over the $V_{ds}$ range, indicating that carrier injection across the junction is effectively quenched by gate-induced depletion.

We first recall the equilibrium band alignment and then consider how the applied gate voltage and bias voltage modify carrier injection at the junction. Figure 2a shows a schematic of the equilibrium band alignment in the Te/MoS$_2$ heterojunction at zero gate voltage. In this configuration, the p-type Te and n-type MoS$_2$ result in hole accumulation in Te and electron accumulation in MoS$_2$. At the interface, the redistribution of free carriers forms a depletion region (indicated by gray dashed lines), where band bending gives rise to a built-in electric field. Previous reports determined the Te and MoS$_2$ work functions ($\Phi_{Te}$ ≈ 4.75 eV and $\Phi_{MoS_2}$ ≈ 4.3 eV) by UPS/KPFM[36,41]. Accordingly, the band diagram in Fig. 2a is constructed using these quantitative data and provides a quantitative basis for the equilibrium band alignment. Importantly, the large valence band offset at the Te/MoS$_2$ interface serves as an effective energy barrier, preventing hole leakage from Te into MoS$_2$.

To rationalize the gate-dependent band diagrams in Fig. 2d, it is important to consider how the back gate voltage couples to the Te and MoS$_2$ layers in this vertical geometry. The Te/MoS$_2$ heterojunction and the back silicon electrode form a capacitor. Because the MoS$_2$ layer is relatively thin and lies directly on the gate dielectric outside Te, the back-gate voltage primarily modulates the carrier density in the edge regions of MoS$_2$ rather than in the region where MoS$_2$ overlies Te. In this geometry, the gate field effectively penetrates the full thickness of the ~17 nm MoS$_2$ channel and tunes its Fermi level, while the much thicker Te is strongly screened and remains only weakly affected by the back gate. This diagram of the gating effect is consistent with the EL spatial map shown in Fig. 1d. The EL-active region extends across the Te/MoS$_2$ overlap area. Although the back gate couples most efficiently to the MoS$_2$ regions that are not covered by Te, carriers can be efficiently injected into the whole overlap area with Te, which contributes to the EL emission from the overlapped area. Consequently, back-gating provides an efficient handle to tune the effective band alignment and injection across the Te/MoS$_2$ interface and thus the EL output from the overlap region.

For $V_g$ = 0 V and +30 V, a finite reverse-bias current persists at sufficiently negative $V_{ds}$, which can be attributed to band-to-band tunneling. As illustrated in Fig. 2d(i) and 2d(ii), the combined effect of the drain–source bias $V_{ds}$ and the back-gate voltage $V_g$ bends the bands such that the conduction band minimum of MoS$_2$ lies below the valence band maximum of Te, resulting in a type-II band configuration. When the magnitudes of $V_{ds}$ are fixed in Fig. 2d, positive $V_g$ shifts the MoS$_2$ conduction band downward in energy relative to Te and increases the overlap between the MoS$_2$ conduction band and Te valence band. This effect enhances the tunneling probability and produces a larger reverse current at $V_g$ > 0 than at $V_g$ = 0, which is consistent with the transport data shown in Fig. 2b. In contrast, for $V_g$ < 0 (Fig. 2d(iii)), the MoS$_2$ conduction band is shifted upward in energy relative to the Te valence-band edge, so the band-to-band tunneling window effectively closes, and the reverse current is strongly suppressed. The transfer characteristics ($I_{ds}$- $V_g$) measured at $V_{ds}$ = 4 V confirm that the device exhibits pronounced rectifying behavior, with the current strongly dependent on gate-induced carrier modulation. The threshold voltage ($V_{th}$) of the MoS$_2$ FET was determined to be approximately -7 V. This value was extracted from the logarithmic $I_{ds}$–$V_g$ curve transfer curve (Fig. 2c) as the gate voltage at which the drain current departs from the subthreshold regime and begins to increase rapidly. These results demonstrate that the Te/MoS$_2$ heterostructure operates as a robust and gate-tunable p–n junction, with rectifying behavior and suppressed leakage arising from the built-in potential at the interface. These characteristics provide a consistent electronic framework for understanding the bias- and gate-dependent electroluminescence discussed later.

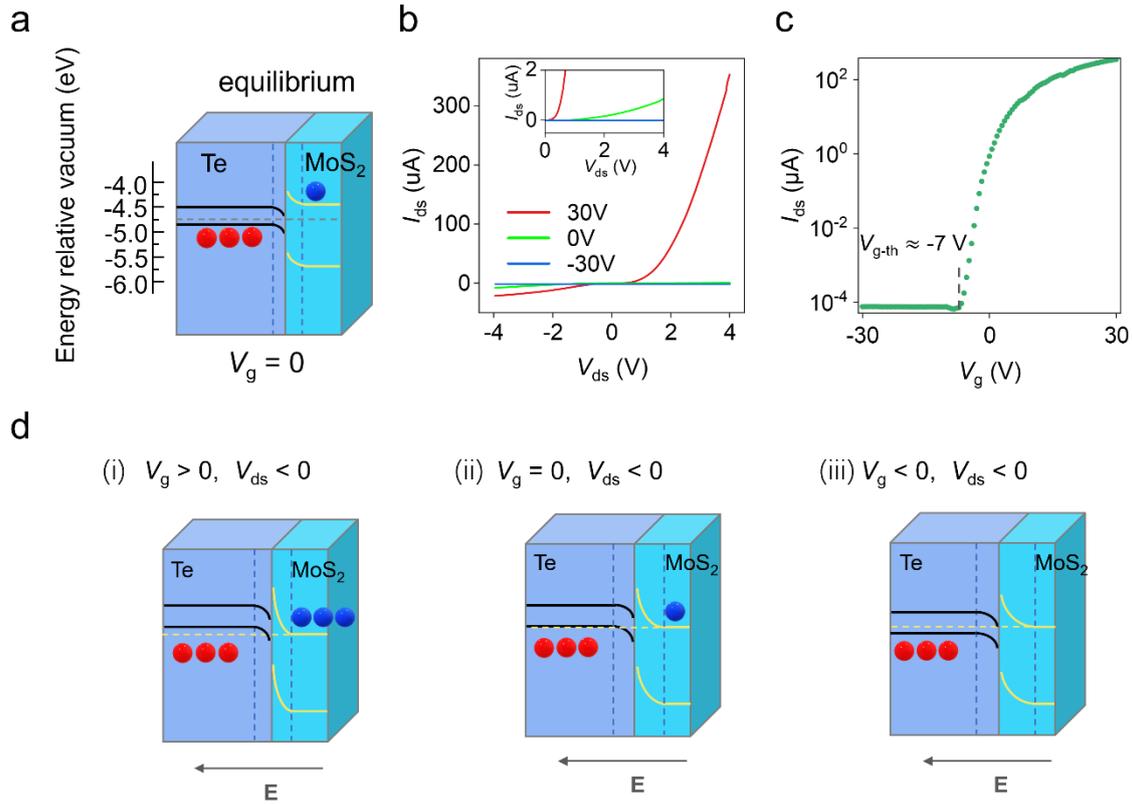

**Fig. 2 | Band alignment and electrical transport of the Te/MoS₂ heterojunction. a,** Qualitative equilibrium band alignment of the Te/MoS$_2$ heterojunction at $V_g$ = 0 V, with the depletion region marked by gray dashed lines. **b,** Output characteristics ($I_{ds}$-$V_{ds}$) of the heterostructure measured at 25 K for back-gate voltages $V_g$ = 30 V, 0 V, and −30 V, showing gate-tunable rectification behavior. The inset provides a magnified view of the low-current region. **c,** Transfer characteristics ($I_{ds}$-$V_{ds}$) recorded at $V_{ds}$ = 4 V. **d,** Schematic band alignment of the Te/MoS$_2$ heterojunction under reverse bias ($V_{ds}$ < 0) for three representative gate voltages: (i) $V_g$ > 0, (ii) $V_g$ = 0, and (iii) $V_g$ < 0. The yellow dashed line indicates the position of the MoS$_2$ conduction band edge.

**Mid-infrared electroluminescence and bias-dependent spectra of Te/MoS₂ heterojunctions**

We next discuss the electroluminescence and underlying recombination mechanisms under different bias conditions. To directly correlate this band-alignment diagram with light emission, we investigate how the mid-infrared EL depends on the applied source–drain bias at a fixed gate voltage. Bias-dependent EL spectra (Fig. 3b) are obtained by sweeping the source-drain voltage ($V_{ds}$) from -4 V to 4 V while keeping the gate voltage fixed at $V_g$ = 20 V. EL emission is observed only when the forward bias exceeds an EL turn-on voltage ($V_{on}$). We define $V_{on}$ as the minimum $V_{ds}$ at which the integrated EL intensity clearly exceeds the background signal. $V_{on}$ is not a fixed constant but varies systematically with gate voltage. At $V_g$ = 20 V, the spectra shown in Fig. 3b indicate that $V_{on}$ > 2 V and $V_{on}$ is accurately determined to be below 1.7 V using the higher sensitivity measurement described in supplementary Information Note 7. Notably, the optical turn-on voltage ($V_{on}$ ~1.7 V) is larger than both the Te bandgap scale ($E_g/q$ ~0.35 V) and the estimated built-in potential (~0.46 V). This is expected because $V_{on}$ reflects the bias required to reach sufficient interlayer injection and quasi-Fermi level splitting in the overlap region for detectable radiative recombination, together with additional voltage drops from contact/access series resistance. Near the onset, trap filling and defect-assisted nonradiative recombination can further delay the emergence of EL to larger voltage. For $V_{ds}$ above $V_{on}$, the EL intensity increases steadily when the applied bias increases further. The EL peak position remains essentially unchanged under different carrier injection levels, indicating a stable underlying radiative transition. The linewidth shows a weak narrowing trend with increasing bias (Supplementary Note 12). As the emission becomes increasingly dominated by band-edge recombination, the relative contribution from the low-energy-side emission decreases, leading to a slight reduction in the full width at half maximum. The low-energy in-gap emission may originate from shallow impurity states, the Franz–Keldysh effect induced by the built-in electric field, and excitonic emission. Figure 3a illustrates the band alignment under forward bias: the applied voltage flattens the electron barrier on the $MoS_2$ side and thereby lowers the effective electron injection barrier; at the same time, the large valence band offset at the Te/$MoS_2$ interface maintains a high barrier for holes in Te. These two conditions together ensure that radiative recombination is spatially localized within the Te layer, which is responsible for the observed mid-infrared emission. The spatial profiles of the quasi-Fermi levels under forward bias lend further support to this interpretation. Specifically, the electron quasi-Fermi level in $MoS_2$ exhibits a gradual downward slope as it approaches the interface, confirming efficient electron injection into Te. In contrast, the hole quasi-Fermi level in Te rises steeply near the boundary, forming a sharp gradient that signifies strong hole confinement and suppressed injection into $MoS_2$. Under reverse bias, as shown in Fig. 2d(i), the applied voltage increases the built-in potential, thereby steepening the quasi-Fermi level gradients on both sides of the heterojunction. This enhancement strengthens carrier separation within the depletion region and effectively suppresses both electron and hole injection. As a result, radiative recombination is inhibited, leading to suppressed electroluminescence under reverse bias. These quasi-Fermi level profiles qualitatively explain the bias-dependent switch of EL emission. Long-term operational stability data, showing that the device maintains robust mid-infrared electroluminescence after extended ambient storage without any encapsulation, are provided in Supplementary Fig. S7.

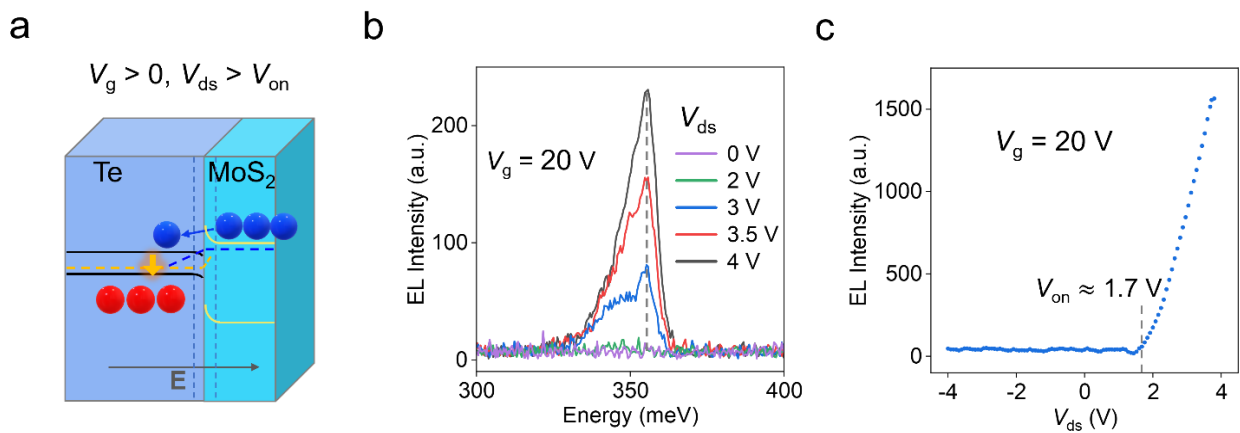

**Fig. 3 | Band alignment and bias-dependent electroluminescence of the Te/$MoS_2$ heterojunction. a**, $V_{ds}$ > $V_{on}$ ($V_g$ =20 V), illustrating the light-emission process. Blue and orange

dashed lines denote quasi-Fermi levels for electrons and holes, respectively. **b**, Bias-dependent EL spectra at $V_g = 20$ V (25 K). **c**, Integrated EL intensity as a function of bias voltage.

**Gate-voltage tunable electroluminescence of Te/MoS$_2$**

Having identified the bias window over which efficient mid-infrared EL occurs, we next explored how electrostatic gating of the MoS$_2$ channel modulates the EL intensity at fixed forward bias. Since EL emission in the Te/MoS$_2$ heterostructure arises from radiative recombination between electrons injected from MoS$_2$ and holes localized in Te, the EL intensity is highly sensitive to the doping level of MoS$_2$. In particular, gate tuning of the doping level of MoS$_2$ modulates the interfacial band alignment, thereby influencing the efficiency of carrier injection into the Te layer and ultimately tuning the light emission intensity. Since the back gate has little effect on the quasi-ohmic Te/Pd contacts (Fig. S2), its modulation of the doping level in MoS$_2$ can slightly affect the built-in potential of MoS$_2$/Cr contact, and further modulates the effective voltage applied on the MoS$_2$/Te heterojunction. However, because Schottky barrier of MoS$_2$/Cr contact is only tens of meV, the effect of the built-in potential changes by the doping is expected to be limited, providing a minor correction compared with the heterojunction band alignment that governs the gate-tunable EL and is therefore omitted from the schematic band diagrams for clarity[42]. To elucidate this effect, we measured the EL response as a function of back-gate voltage $V_g$ at $V_{ds} = 4$ V. Figure 4a displays the EL emission energy map under a fixed lateral bias ($V_{ds} = 4$ V), revealing a nonmonotonic intensity profile: the EL emission shows a maximum at a critical gate voltage ($V_{g\text{-crit}}$) of approximately 20 V.

This trend results from competition between enhanced electron accumulation and the formation of a gate-induced effective interfacial injection barrier. The underlying mechanism is illustrated by the band diagrams in Fig. 4b and 4c. Specifically, Fig. 4b highlights two representative gating conditions that define the onset and maximum of EL emission, while Fig. 4c traces the full evolution of the equilibrium band structure across the entire gating range. Figure 4b(i) corresponds to the critical gate voltage ($V_{g\text{-crit}} = 20$ V), where the EL emission is maximized because enhanced electron accumulation in MoS$_2$ improves the electron supply while the gate-induced interfacial injection barrier increases. The competition between these two effects reaches an optimum balance at $V_{g\text{-crit}}$. The quasi-Fermi-level profile in Fig. 4b(i) indicates that the MoS$_2$ channel resistance is reduced at this gate bias due to a high carrier density, so that under the same applied $V_{ds}$, a larger fraction of the voltage can drop across the junction region. Consistently, the electron quasi-Fermi level in MoS$_2$ varies more gently toward the interface, which reflects a smaller series voltage drop in the MoS$_2$ access region and thus more favorable injection conditions into Te. Meanwhile, the large valence-band offset at the Te/MoS$_2$ interface blocks hole injection from Te into MoS$_2$. Consistently, the hole quasi-Fermi level on the Te side varies sharply near the interface, indicating strong hole confinement. Figure 4b(ii) represents the threshold voltage ($V_{g\text{-th}} = 0$ V), where the gate-induced electron population in MoS$_2$ first becomes sufficient to enable injection into Te under forward bias. In this case, the electron quasi-Fermi level exhibits a steeper spatial variation toward the junction, consistent with a larger series drop in MoS$_2$ and injection that is only just initiated.

Figure 4c illustrates the full evolution of the band alignment in the Te/MoS$_2$ heterostructure, complementing the specific gating conditions in Fig. 4a to offer a more comprehensive view. Three distinct gate-controlled regimes are identified:

(i) $V_g < V_{g\text{-th}}$, Fig. 4c(i), (the depleted regime): below the threshold ($V_{g\text{-th}}$), MoS$_2$ is depleted of free electrons. The lack of available carriers quenches EL emission.

(ii) $V_{g\text{-th}} < V_g < V_{g\text{-crit}}$, Fig. 4c(ii), (intermediate regime): as $V_g$ increases above $V_{g\text{-th}}$, electron accumulation in $MoS_2$ increases, improving the electron supply for injection. The gate reshapes the electrostatic band bending near the interface and can increase the effective electron injection barrier. In this gate range, the benefit from increased electron supply dominates over the barrier penalty, so the net electron injection into Te and the EL intensity still increase with $V_g$ and reach a maximum at $V_{g\text{-crit}}$.

(iii) $V_g > V_{g\text{-crit}}$, Fig. 4c(iii), (above 20 V, the high gate voltage regime): a higher gate voltage reshapes the electrostatic potential and band bending near the interface. This increases the effective electron injection barrier for carriers entering Te. The increased barrier suppresses electron injection into Te and becomes the dominant factor. It outweighs the benefit from additional electron accumulation, so the EL intensity no longer increases and decrease beyond $V_{g\text{-crit}}$.

To gain further insight into the gate-dependent EL behavior, we investigated whether the underlying radiative transitions responsible for the observed EL remain constant across different gate voltages. As shown in Fig. 4d, the emission peak position remains fixed at 3.5 μm with negligible linewidth variation, indicating a stable recombination pathway governed by band-edge transitions in Te, even as the intensity varies. This confirms that gate modulation primarily affects the carrier injection efficiency rather than the fundamental nature of the recombination mechanism in Te. Together with the bias-invariant spectrum shown in Fig. 3b, these results underscore the robustness of Te band-edge transitions against electrical perturbations. In addition, the EL polarization state remains essentially unchanged during back-gate sweeps at a fixed drain bias. As shown in Supplementary Fig. S9, the extracted degree of linear polarization (DOP) stays within ~0.58–0.59 over a broad gate range while the EL intensity varies, demonstrating a polarization-locked intensity control enabled by electrostatic tuning. In contrast to materials such as BP, whose emission energy shifts with applied vertical fields[28], the gate bias in our Te/$MoS_2$ heterojunction modulates EL intensity without inducing measurable changes in emission wavelength, demonstrating stable band-edge emission under field control. To further establish the gate dependence of EL, we measured the EL intensity as a function of $V_g$, as shown in Fig. 4e. As the gate voltage is swept from positive toward negative values, the EL intensity decreases rapidly and is effectively quenched. Based on the EL–$V_g$ intensity map shown in Fig. 4e, the EL threshold is more precisely identified as $V_{g\text{-th}} = -2.7$ V. The refined threshold provides a clear reference point for subsequent analysis of the current-dependent EL efficiency (see Fig. 5).

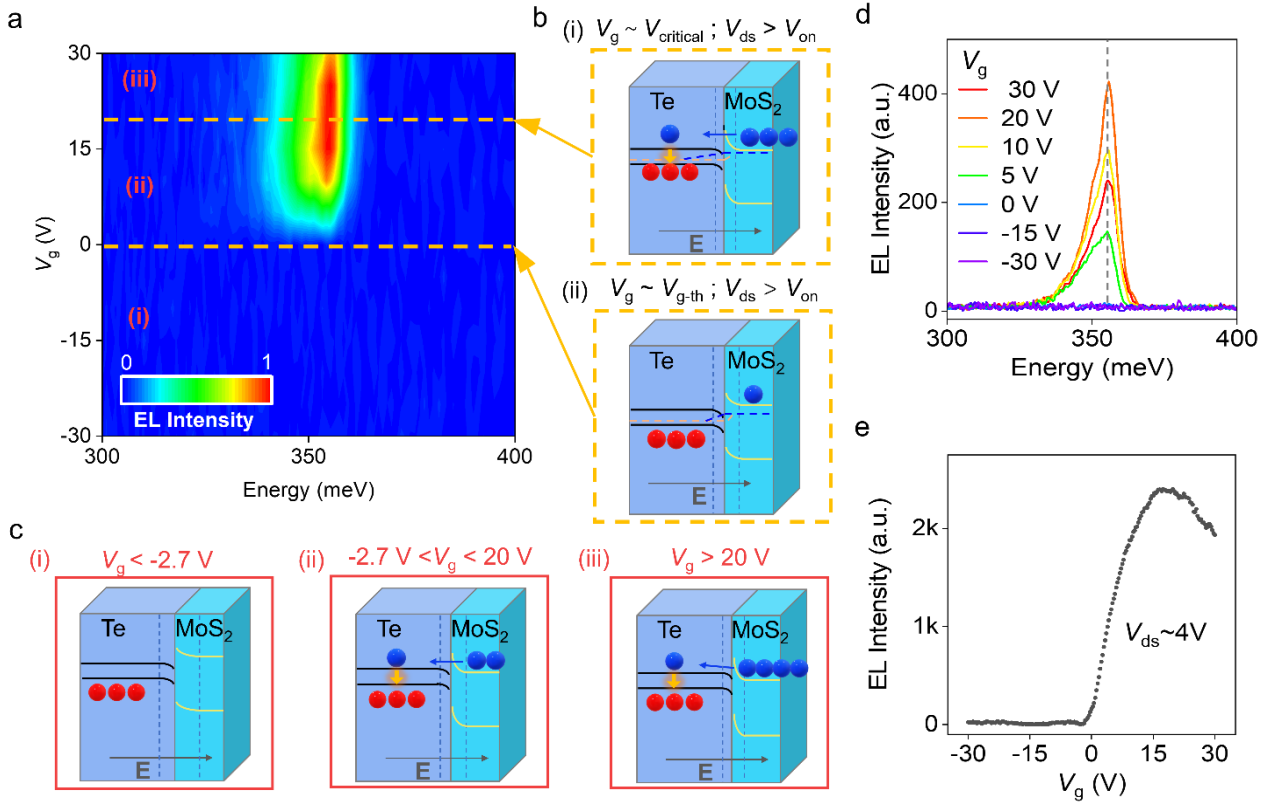

**Fig. 4 | Gate-voltage dependence of Te/MoS₂ electroluminescence. a**, Plot of EL intensity as a function of photon energy and gate voltage $V_g$ under a 4 V bias voltage. **b**, Interfacial energy band alignment diagram of Te/MoS₂ under critical gate voltages and threshold gate voltages, where diagrams in (i) and (ii) correspond to the energy bands of the yellow dashed lines in Fig. 4a. The blue and orange dashed lines in (i) and (ii) indicate the quasi-Fermi levels for electrons and holes, respectively. **c**, Interfacial energy band alignment of Te/MoS₂ across different gate voltage regions: (i) non-radiative states under $V_g < -2.7$ V, (ii) intermediate regime where $V_g < V_{g\text{-crit}}$, and (iii) radiative bands at $V_g > V_{g\text{-crit}}$. **d**, Selected EL spectra at different gate voltages, ranging from 30 V to -30 V, measured at 25 K. **e**, Integrated intensity of the EL emission as a function of gate voltage (with fixed $V_{ds}$=4 V) , measured at 25 K.

**Driving-current dependence of mid-infrared electroluminescence**

To probe how the bias and gate voltage affect the internal quantum efficiency and carrier injection efficiency, we extract the differential power-law exponent $k$ from log–log plots of $I_{EL}$ vs $J_{ds}$ (calculated using a junction area A ~300 μm² via $J_{ds} = I_{ds}/A$) . Following the k-analysis commonly used for LEDs, the integrated EL intensity can be written as the product of the source-drain current term ($I_{ds}$), the injection efficiency ($\eta_{inj}$), and the internal quantum efficiency ($\eta_{IQE}$) [43-45], so that

$$I_{EL} \propto I_{ds} \cdot \eta_{inj} \cdot \eta_{IQE}, \tag{1}$$

In the active-region-limited regime, this leads to

$$k = \frac{d \ln I_{EL}}{d \ln I_{ds}} = 1 + \frac{d\ln \eta_{IQE}}{d\ln I_{ds}} + \frac{d\ln \eta_{inj}}{d\ln I_{ds}} = k_\perp + \frac{d\ln \eta_{inj}}{d\ln I_{ds}}, \tag{2}$$

where $\eta_{inj}$ characterizes the injection efficiency of carriers that can cross the Te/MoS₂ interface barrier and enter the Te recombination zone and $k_\perp$ reflects the competition between radiative and nonradiative processes inside Te. In conventional LED recombination models for LEDs, Shockley–Read–Hall (SRH) defect recombination, band-to-band radiative recombination, and Auger recombination correspond to characteristic values of $k_⊥ ≈ 2$, 1, and 2/3, respectively[46-48]. Experimentally, we sweep $V_{ds}$ at a fixed gate voltage ($V_g = 20$ V, Fig. 5a) and sweep $V_g$ at a fixed bias ($V_{ds} = 4$ V, Fig. 5b) to study the dependence of EL ($I_{EL}$) on the drain current ($I_{ds}$).

Figure 5a shows $I_{EL}$ versus current density $J_{ds}$ on a double-logarithmic scale at $V_g = 20V$, a bias chosen to maximize the EL signal-to-noise ratio. Between the EL onset at $I_{ds} ≈ 0.05$ μA/μm² and $I_{ds} ≈ 0.4$ μA/μm², the dependence is super-quadratic dependence with an effective exponent $k ≈ 2.36$. This large $k$ indicates that defect-assisted SRH recombination in Te remains the dominant recombination channel [47-49]. As $V_{ds}$ increases, the Te/MoS$_2$ barrier is lowered and thinned, increasing the fraction of the total current, $I_{ds}$, injected into Te. In the framework of Eq. (1), this corresponds to a positive $d(\ln η_{inj})/d(\ln I_{ds})$, which drives the measured exponent slightly above the SRH limit. For higher currents, $0.4$ μA/μm² $≲ J_{ds} ≲ 1.3$ μA/μm², the slope decreases to a super-linear $k ≈ 1.29$. In this regime, radiative recombination in Te becomes dominant (~1) [48,49], and the injection efficiency continues to improve with increasing $V_{ds}$, so $k$ remains slightly above 1.

To separate the effects of forward bias and gate control while maintaining a high signal-to-noise ratio, we next fix $V_{ds} = 4$ V and tune $V_g$ (Fig. 5b). When the gate-induced current lies in the range $J_{ds} ≈ 0.03–0.3$ μA/μm², $I_{EL}$ again follows a super-linear power law with $k ≈ 1.89$. This value indicates that defect-related SRH recombination in the active region remains important, so $k_⊥$ is close to the SRH limit (~2)[47-49]. In this regime, increasing $V_g$ at fixed $V_{ds} = 4$ V rapidly increases the MoS$_2$ channel conductance and raises the effective Te/MoS$_2$ junction barrier. A smaller fraction of $I_{ds}$ reaches the radiative zone, and a larger fraction is dissipated via non-radiative or interfacial dark paths outside the EL-active region. In the language of Eq. (1), $η_{inj}$ decreases with $I_{ds}$, and this negative trend lowers the measured $k$ from the SRH-like value. For a higher current $0.3$ μA/μm² $≲ J_{ds} ≲ 1.3$ μA/μm², the slope drops to $k ≈ 0.71$. In this regime, radiative recombination in Te remains relevant[48,49], but the gate-induced redistribution of potential increasingly favors non-radiative/interface loss channels outside the EL-active region, leading to a droop of $η_{inj}$ and driving $k$ below 1. At the largest drive currents ($J_{ds} ≳ 1.3$ μA/μm²), the EL intensity eventually rolls over and decreases with increasing $J_{ds}$, corresponding to an effective negative exponent in the $I_{EL}$-$J_{ds}$ power-law relation.

These trends can be consistently interpreted within a standard LED framework that separates injection efficiency from the internal quantum efficiency of the Te active region. In the low-current-density window (0.03–0.3 μA/μm²), back-gating accumulates electrons in the MoS$_2$ edge/access regions and lowers the effective injection barrier into Te, thereby suppressing the SRH-type losses in Te and increasing $η_{inj}$ with current; together, these effects naturally yield super-linear exponents approaching the SRH-like limit. As $V_g$ increases further, the junction enters a regime in which gate-induced band bending on the MoS$_2$ side becomes substantial (cf. Fig. 2), and a larger fraction of $I_{ds}$ is shunted through the MoS$_2$ rather than recombining in Te. In other words, $η_{inj}$ begins to decrease with current ($d\ln η_{inj}/d\ln I_{ds} < 0$), which by itself can drive the observed sub-linear behavior ($k ≈ 0.71$) even if radiative recombination remains important in Te. At the highest drive currents ($J_{ds} ≳ 1.3$ μA/μm²), the slope $k$ dips and becomes negative. As the carrier density increases, high-order non-radiative processes such as Auger recombination become significant, and self-heating further enhances non-radiative loss. The gate-induced barrier and dark path leakage continue to reduce the fraction of $I_{ds}$ that reaches the EL-active region, so $k_⊥$ tends toward the Auger value (~2/3)[47,50,51], while $d\ln η_{inj}/d\ln I_{ds}$ remains negative; together, these factors drive the overall $k$ below zero.

To quantify the device efficiency, we evaluate the external quantum efficiency (EQE) of the Te/MoS$_2$ heterojunction LED. EQE is defined as the ratio of photons emitted into the far field to electrons passing through the device, i.e., EQE = $Φ_{ph}/(I_{ds}/q)$. The photon flux $Φ_{ph}$ is obtained from the measured EL intensity after correcting for the calibrated throughput of the collection optics and detector, as detailed in Supporting Information Note 6. Figure 5c shows the EQE as a function of drive current $I_{ds}$ for the same devices, where the red curve corresponds to sweeping $V_{ds}$ at fixed $V_g$ = 20 V and the blue curve corresponds to sweeping $V_g$ at fixed $V_{ds}$ = 4 V. Within our IQE–injection-efficiency framework, the integrated EL intensity scales as $I_{EL} ∝ I_{ds} η_{inj} η_{IQE}$, which yields:

$$d\ln(\text{EQE})/d\ln(I_{ds}) = k - 1 \qquad (3)$$

Thus, $k > 1$ indicates that the EQE increases with current, $k ≈ 1$ marks the vicinity of the EQE maximum, and $k < 1$ signals the onset of efficiency droop. For the forward-bias sweep at fixed $V_g$ = 20 V (red curve), the EQE continues to increase with $J_{ds}$ over the measured range, as expected from the super-linear exponents k>1 (Fig. 5a). Under these conditions, a maximum EQE of ~0.10% is obtained. In contrast, for the gate sweep at fixed $V_{ds}$ = 4 V (blue curve), the EQE initially rises steeply with increasing $J_{ds}$, reaches a peak (EQE ≈ 0.32%) near the current where $k ≈ 1$, and then decreases to ~0.10% as $k$ drops below unity and eventually becomes negative (Fig. 5b). The shapes and peak positions of the EQE–$J_{ds}$ trace closely mirror the segmented k-value behavior discussed above, providing a consistency check of the combined IQE–injection-efficiency picture. In this three-terminal heterojunction, the drain bias and the gate voltage act as two independent knobs for controlling the EL output. The drain bias mainly sets the overall forward-bias drive and thus the injection level, which naturally modulates the EL intensity. In contrast, the back gate electrostatically tunes the MoS$_2$ carrier density, altering the resistance and, consequently, the injection current. However, the distinct EQE-$J_{ds}$ curves for $V_{ds}$-sweep and $V_g$-sweep in Fig. 5c indicate that the back gate provides additional control over the device: it can redistribute the potential drop near the Te/MoS$_2$ junction, thereby modifying the effective interfacial injection condition and the radiative probability per injected carrier. As a result, gating can shift the device into a higher-efficiency operating window and enhance the EQE without requiring a proportional increase in $J_{ds}$. Notably, a peak EQE of ~0.32% is achieved in a simple planar device geometry without any optical cavity, reflective back contact, or waveguide-based light extraction. In light of recent reports on mid-infrared BP-based LEDs employing resonant cavities and reflective electrodes[23,52], we anticipate that similar optical and interfacial engineering could further enhance the EQE of Te/MoS$_2$ heterojunction LEDs.

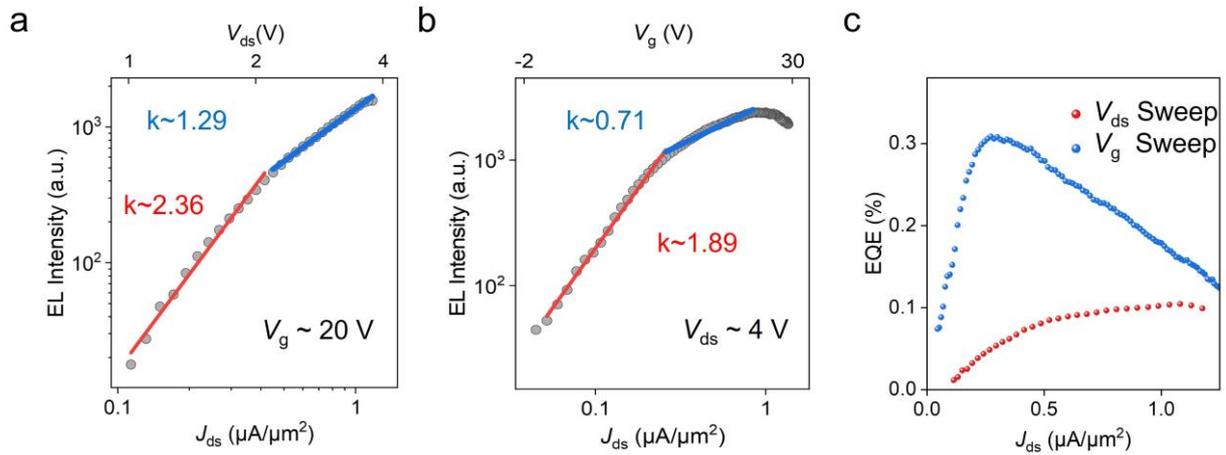

**Fig. 5 | Current-dependent electroluminescence in Te/MoS$_2$ heterojunctions.** Log–log plots of integrated EL intensity versus drain–source current density ($J_{ds}$), with red and blue solid lines indicating power-law fits to different current regimes. **a**, Variation under different source–drain bias, showing distinct low- and high-current-density scaling. **b**, Variation under different gate voltages, exhibiting segmented power-law behavior with fitted exponents reported in the text. **c**, External quantum efficiency (EQE) versus $J_{ds}$; red and blue curves correspond to the bias sweep and gate sweep configurations in (a) and (b), respectively.

## Conclusion

In summary, we demonstrate a gate-tunable Te/MoS$_2$ van der Waals LED emitting at ~3.5 μm, a technologically important mid-IR wavelength, with linearly polarized output and spectra that remain invariant to bias and gate at 25 K. The device architecture enables electrostatic control of band alignment and injection efficiency, while a segmented k-slope analysis identifies injection-efficiency droop as the principal limitation at high drive. These results position the Te/MoS$_2$ heterostructure as a polarization-sensitive MIR source compatible with heterogeneous integration, with immediate implications for on-chip sensing, spectroscopy, and communications in the 3–5 μm window. In particular, polarization-sensitive MIR emission and the potential for integration with mature photonics platforms open opportunities for compact, chip-scale spectrometers, trace-g gas sensing, environmental monitoring, and polarized MIR communication systems that leverage atmospheric transmission windows.

Looking forward, concrete routes to higher brightness and elevated operating temperature include contact and barrier engineering to boost injection while suppressing leakage, refined thermal management or pulsed operation to mitigate roll-over, and photonic confinement strategies (such as microcavities, waveguide-coupled resonators, or photonic crystals) to enhance out-coupling. Realization on CMOS-friendly substrates and co-integration with Si or other MIR photonics platforms would enable lab-on-a-chip sensing, portable MIR spectrometers, and field-deployable LIDAR or free-space links. Beyond this system, the segmented k-slope framework provides a generalizable toolkit for vdW emitters and can guide the rational design of reconfigurable, on-chip MIR optoelectronics, including multi-color or tunable polarization-enabled sources.

## Methods

### Sample Preparation

Te nanoflakes were grown through a hydrothermal method[33,34]. Initially, 3 g of polyvinylpyrrolidone (PVP, molecular weight = 58000) was dissolved in 32 mL of deionized (DI) water. Subsequently, 92 mg of $Na_2TeO_3$ was put into the PVP solution with continuous stirring. Next, 3.32 mL of ammonium hydroxide solution (25–28%, wt/wt%) and 1.68 mL of hydrazine hydrate (80%, wt/wt%) were added to the mixture. After five minutes of magnetic stirring, the solution was transferred into a 50 mL Teflon-lined stainless-steel autoclave and maintained at 180°C for 10 hours. The resulting product was washed thoroughly with deionized water to remove any residual ions.

For device fabrication, the Te nanoflakes were redispersed in ethyl alcohol and drop-casted onto a 285-nm $SiO_2$/Si substrate. Multilayer $MoS_2$ flakes were exfoliated using polydimethylsiloxane (PDMS) and transferred onto selected Te flakes via a dry transfer method. Electrode fabrication was carried out in two sequential electron beam lithography (EBL) steps. First, electrodes were patterned on the Te flake, followed by electron beam evaporation of 20 nm Pd and 100 nm Au (Pd/Au). In the second EBL step, electrodes were defined on the $MoS_2$ region, and a 20-nm Cr and 100-nm Au (Cr/Au) metal stack was deposited using the same electron beam evaporation process. Each lithography and deposition step was followed by a standard lift-off process in acetone.

**Mid-infrared EL Measurement**

EL measurements were performed using a self-built micro-EL setup. The sample was mounted in an open-loop cryogenic thermostat providing a vacuum environment and temperature control from 25 K to 300 K via a liquid helium flow. The cryostat was fixed onto a PDV XY-50 motorized XY scanning stage to enable spatially resolved measurements. A ×40 reflective objective lens with a numerical aperture of 0.5 was used to collect the EL emission, which was directed into a spectrometer (Princeton Instruments SP2500i) equipped with a 150 g/mm grating. The signal was detected using a liquid-nitrogen-cooled single-channel InSb photodetector. To improve the signal-to-noise ratio, the EL signal was modulated by an optical chopper at 521 Hz and demodulated using a lock-in amplifier phase-locked to the chopper reference. The optical chopper was placed in the collection path. During the measurements, the electrical bias was applied using a Keithley 2612B dual-channel source meter, with one channel supplying the back-gate voltage and the other applying the drain-source bias.

**PL and Raman Measurement**

For photoluminescence (PL) measurements, the same low-temperature cryogenic setup, motorized XY stage, and lock-in detection scheme described for EL were used. A 1064 nm continuous-wave laser was employed as the excitation source. The PL signal was collected by the same objective and detected using the same InSb photodetector. For Raman spectroscopy, a 532-nm laser was used for excitation, and the scattered signal was collected by the same spectrometer with the same grating and detected using a CCD camera.

## Acknowledgments
This project was supported by the National Natural Science Foundation of China (Grant Nos. 62325401 [D.S.] and 12034001 [D.S.]) and the National Key Research and Development Program of China (Grant Nos. 2021YFA1400100 [D.S.]). The authors would also like to acknowledge the National Key Research and Development Program of China (Grant Nos. 2022YFA1204300 [A.L.P.]), the National Natural Science Foundation of China (Grant Nos., 12034003 [J.L.C.], 52221001 [A.L.P.], 62090035 [A.L.P.], 62250065 [D.S.], and 62227822 [D.S.]), and the Open Fund of State Key Laboratory of Infrared Physics (Grant No. SITP-NLIST-ZD-2023-02 [D.S.]).

## Author contributions
D.S. conceived the idea and supervised the project. Z.Z. synthesized the Te nanoflakes under the supervision of L.L. and C.G.Z.; S.Y.W. and D.L.L. performed the optical and electrical measurements with the help of M.Y.Q., Y.C.C., J.S., and S.L.C. under the supervision of A.L.P., J.L.C. and D.S.; W.S.Y, J.L.C. and D.S. wrote the manuscript with input and feedback from all authors.

## Competing interests
The authors declare no competing interests.

# Supplementary Materials for

# Gate-Tunable Mid-Infrared Electroluminescence from Te/MoS$_2$ p–n Heterojunctions

Shiyu Wang[1], Delang Liang[1,2], Zhi Zheng[3], Mingyang Qin[1], Yuchun Chen[1], Jie Sheng[1], Shula Chen[2], Lin Li[3] Changgan Zeng[3], Anlian Pan[2,7][†], Jinluo Cheng[4,8][†], and Dong Sun[1,5,6,8][†]

[1]International Center for Quantum Materials, School of Physics, Peking University, Beijing, China.
[2]Key Laboratory for Micro-Nano Physics and Technology of Hunan Province, Hunan Institute of Optoelectronic Integration, College of Materials Science and Engineering, Hunan University, Changsha, China.
[3]CAS Key Laboratory of Strongly Coupled Quantum Matter Physics, and Department of Physics, University of Science and Technology of China, Hefei, Anhui, China.
[4]GPL Photonics Laboratory, State Key Laboratory of Luminescence Science and Applications, Changchun Institute of Optics Fine Mechanics and Physics, Chinese Academy of Sciences, Changchun, China.
[5]Collaborative Innovation Center of Quantum Matter, Beijing, China.
[6]Frontiers Science Center for Nano-optoelectronics, School of Physics, Peking University, Beijing, China.
[7]School of Physics and Electronics, Hunan Normal University, Changsha, China.
[8]School of Physics and Laboratory of Zhongyuan Light, Zhengzhou University, Zhengzhou, China.

[†]Email: sundong@pku.edu.cn; jlcheng@ciomp.ac.cn; anlian.pan@hnu.edu.cn;

List of contents:

Supplementary figures:

**Supplementary Note 1. Optical image and AFM thickness mapping**

Figure S1a identifies the device layout and the AFM scan window (yellow dashed box); the inset shows the AFM scan from which two representative line profiles were taken along the cyan and green dashed lines. Figure S1b plots the corresponding AFM height traces, with the upper (cyan) profile crossing the $MoS_2$ terrace edge and the lower (purple) profile crossing the Te step, yielding thicknesses of 17 nm for $MoS_2$ and 234 nm for Te, as quoted in the main text. The AFM provides a geometric reference for the optical and electrical measurements reported in the manuscript. The Te flake used in this device is 234 nm thick. We selected Te flakes by screening the PL response of multiple candidates. By PL characterization of multiple Te flakes from the same hydrothermal batch, we found that the thinnest flakes often show less robust morphology and lower emission yield. Therefore, we chose Te flakes with sufficiently uniform morphology to ensure stable junction formation and reliable EL operation. For the n-type injecting layer, we use relatively thick $MoS_2$ (~17 nm) rather than a monolayer or thin layer to ensure robust electronic transport and adequate current drive under practical contact configurations. This choice helps reduce the effective series resistance and enables reproducible EL turn-on. On the other hand, electrostatic screening generally becomes stronger as the thickness increases; therefore, a relatively thin $MoS_2$ layer is required to preserve effective gate tunability of the device.

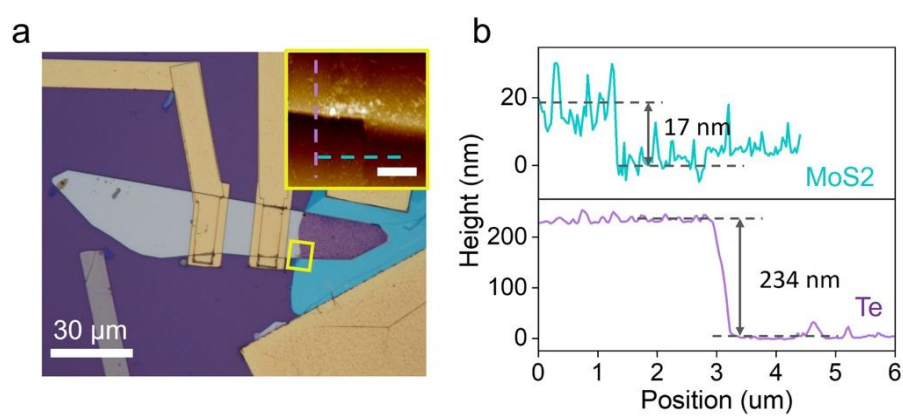

**Figure S1. Thickness of the $MoS_2$/Te device. a,** Optical image of the heterostructure; the yellow rectangle marks the region imaged by AFM (inset). In the AFM inset, cyan and purple dashed guide lines indicate the locations of the line traces plotted in b. **b,** AFM height profiles (height vs position) scanning along the dashed lines marked in panel a (inset): the upper cyan trace shows the $MoS_2$ thickness, and the lower purple trace shows the Te thickness. The step heights are consistent with the values given in the main text. Scale bars: 30 μm (a), 1 μm (AFM inset).

**Supplementary Note 2. Low-temperature transport of Te and Te/$MoS_2$**

Figure S2 presents the 25 K transport measurements; the Te output characteristics, Te transfer characteristics, and Te/MoS$_2$ heterojunction transfer characteristics are presented in panels a–c, respectively. For the isolated Te flake (panels a–b), a nearly linear small-bias $I_{ds} - V_{ds}$ response was observed (Ohmic contacts), and an effective small-bias resistance of ~ 2.3 kΩ was extracted; the $I_{ds} - V_g$ curve decreased monotonically with $V_g$ consistent with p-type conduction. Its transfer curve exhibited weak back-gate modulation at 25 K, changing from ~ 6 µA at $V_g = -60$ V to 3.7 µA at $V_g = +60$ V, consistent with strong electrostatic screening in the Te flake, and consequently, the gate modulation was very limited.

For the Te/MoS$_2$ heterojunction (panel c), transfer curves acquired at $V_{ds} = +4, -2$, and $-4$ V revealed gate-dependent turn-on in forward bias, whereas the reverse-bias branch remains suppressed. A small reverse current was also observed at $V_{ds} = -4$ V near $V_g = +30$ V. This asymmetry is attributed to back-gating that primarily tunes the MoS$_2$ Fermi level (thin and strongly coupled), whereas the much thicker Te is electrostatically screened; as a result, the interfacial barrier and forward injection are modulated, while the basic rectification remains unchanged. Because a single heterojunction transfer trace at $V_{ds} = +4$ V is already presented in the main text, additional transfer curves ($V_{ds} = +2$ V, $-4$ V) are included here to complete the bias–gate matrix while avoiding redundancy in the main figures.

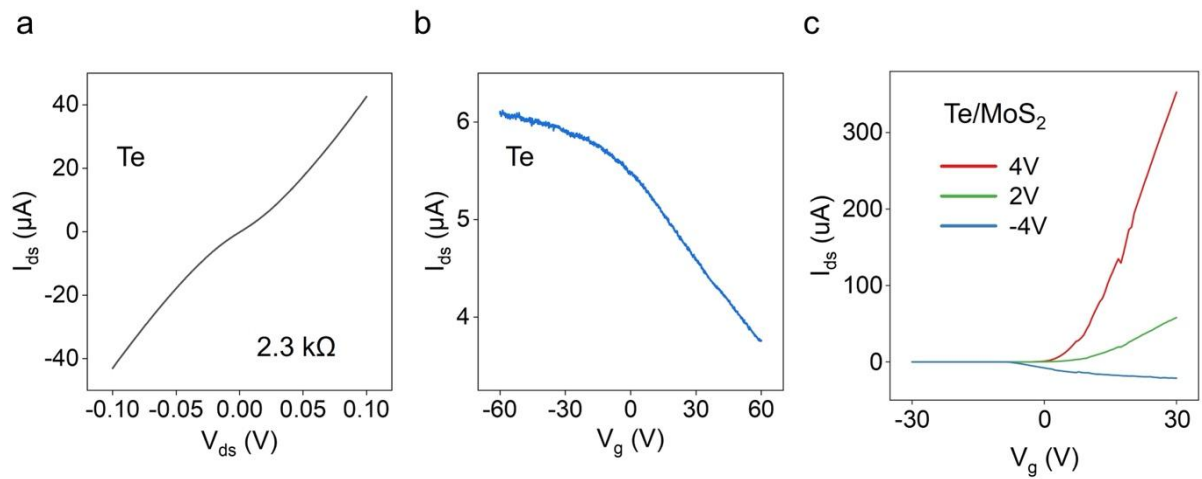

**Figure S2. Electrical characterization at 25 K. a,** Te output characteristics, $I_{ds} - V_{ds}$. **b,** Te transfer characteristics, $I_{ds} - V_g$. **c,** Te/MoS$_2$ heterojunction transfer characteristics at $V_{ds} = +4, +2$, and $-4$ V.

**Supplementary Note 3. Power-dependent photoluminescence**

Figure S3a shows PL spectra from the junction region at 25 K while sweeping the excitation power from 0.53 mW to 49.5 mW. The peak position remains essentially unchanged. The integrated PL versus power (Figure S3b) follows segmented power laws $I_{PL} \propto P^k$ with exponents $k_1 = 1.9$ (low power, super-linear), and $k_2 = 1.1$ (high power). The emission arises from Te in the junction region, while MoS$_2$ does not contribute to the observed emission in the energy regions in these measurements.

The super-linear behavior ($k \sim 1.9$) in the low-power region originates from trap filling (SRH) in Te. With increasing power, laser-induced heating and Auger losses reduce the incremental PL gain, so the slope k drops ($k \sim 1.1$). (*34*) Integrated PL was obtained using the same spectral window for all power levels, and k was determined from linear fits to log–log $I_{PL}$–Power plots.

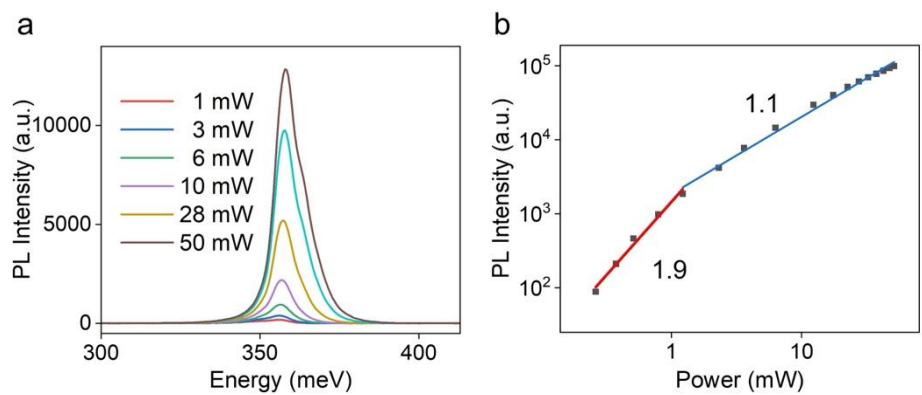

**Figure S3. Power-dependent PL and log–log analysis. a,** PL spectra at 25 K from 0.53 mW to 49.5 mW excitation. **b,** Log–log plot of integrated PL intensity versus power with piecewise linear fits.

**Supplementary Note 4. Temperature dependence of gate-modulated electroluminescence**

Figure S4a compares the EL integrated intensity versus $V_g$ at 25 K, 40 K, and 80 K for the same device and optical path. These data demonstrate that the device continues to exhibit electroluminescence up to 80 K, with the absolute intensity decreasing as temperature increases. At each temperature, the EL always rises and then falls when $V_g$ increases; the absolute intensity follows 25 K > 40 K > 80 K. The common trend is attributed to gate-tunable band alignment at the Te/MoS$_2$ interface: back-gate modulation of the MoS$_2$ Fermi level adjusts the interfacial injection barrier, while higher temperature enhances nonradiative loss and lowers the absolute intensity. Although the MoS$_2$ band gap shifts with temperature (Varshni-type behavior), the overall EL – $V_g$ trend remains unchanged across 25–80 K, as shown in Figure S4b. This interpretation is consistent with the non-monotonic EL – $V_g$ response in the main text, which shows back-gate control of the Te/MoS$_2$ band alignment.

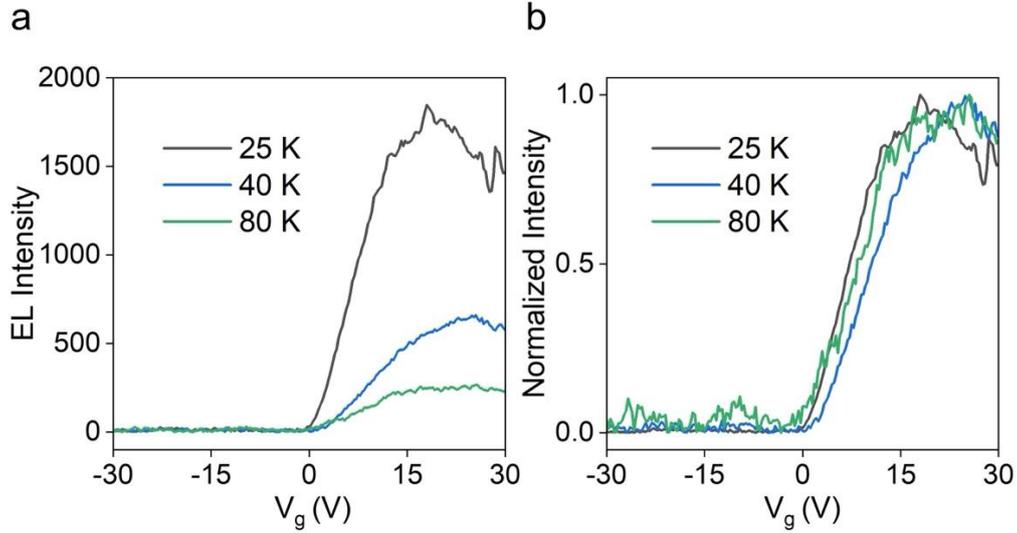

**Figure S4. Gate-voltage dependence of EL at 25 K, 40 K, and 80 K. a,** EL integrated intensity versus $V_g$ at 25 K, 40 K, and 80 K (same device/optical path). **b,** Normalized EL – $V_g$ curves (25 K, 40 K, 80 K).

**Supplementary Note 5. External quantum efficiency (EQE) calibration and calculation**

In the main text, the external quantum efficiency of the Te/MoS$_2$ p–n heterojunction LED is defined as the ratio between the number of photons emitted into the far field and the number of electrons passing through the device:

$$\eta_{\text{EQE}} = \Phi_{\text{ph}}/(I_{\text{ds}}/q),$$

where $\Phi_{\text{ph}}$ is the photon flux emitted into the collection solid angle, $I_{\text{ds}}$ is the measured source–drain current, and q is the elementary charge. The main task is therefore to convert the measured EL signal into an absolute photon flux.

To quantify $\Phi_{\text{ph}}$, we use a calibrated mid-infrared optical setup. The EL emission from the Te/MoS$_2$ junction is collected by a reflective objective, passes through a set of mirrors, beam splitters, and relay optics, is dispersed by a monochromator, and is finally detected by a liquid-nitrogen-cooled InSb photodetector. For each bias condition, we record the spectrally integrated EL signal $I_R$, EL (background-subtracted and integrated over the full EL peak) from the InSb detector.

The overall throughput of the collection optics and detection chain is calibrated using a quantum cascade laser (QCL) and a Lambertian reflectance standard. A QCL emitting at a reference wavelength $\lambda_{\text{ref}}$ is focused onto a diffuse reflector, which approximates a Lambertian source. The QCL output power at the sample plane, $P_{\text{ref}}$, is first measured by a calibrated thermal power sensor placed at the focal plane of the reflective objective. Under identical optical alignment, the QCL light reflected from the Lambertian standard is then directed through the full collection optics and monochromator to the InSb detector, and the corresponding spectrally integrated signal $I_{R,\text{ref}}$ is recorded. Because the QCL output can exceed the detector's linear range, we placed a calibrated attenuator stack directly in front of the LN2-cooled InSb detector during QCL calibration, reducing the detected QCL signal to the same level as in EL measurements. The total transmission at $\lambda_{\text{ref}}$ is denoted as T. With the attenuator stack in place during calibration, the ratio $(P_{\text{ref}} \times T)/I_{\text{R,ref}}$ provides the conversion factor from detector signal to optical power at $\lambda_{\text{ref}}$ for our specific collection geometry.

Because the calibration wavelength $\lambda_{\text{ref}}$ (4.05 μm) is close to the EL peak wavelength $\lambda_{\text{EL}}$ (≈3.5 μm) and the overall response of the optics and detector is nearly flat in this spectral window, we neglect the effect due to wavelength dependence and directly use the same conversion factor for the EL analysis. Under this approximation, the emitted optical power associated with the EL signal is:

$$P_{\text{EL}} = P_{\text{ref}} \times T \times (I_{\text{R,EL}}/I_{\text{R,ref}})$$

The emitted photon flux is obtained by dividing this power by the photon energy at the EL peak (or, equivalently, by using the intensity-weighted mean photon energy extracted from the EL spectrum):

$$\Phi_{\text{ph}} = P_{\text{EL}}/E_{\text{EL}} = P_{\text{EL}}/(\text{hc}/\lambda_{\text{EL}})$$

where $E_{\text{EL}}$ is the photon energy corresponding to $\lambda_{\text{EL}}$, h is Planck's constant, and c is the speed of light. Combining the above expressions yields the working formula used to extract the EQE from the measured EL signal and drive current:

$$\eta_{\text{EQE}} = \Phi_{\text{ph}}/(I_{\text{ds}}/\text{q}) = [(\text{q}\, P_{\text{ref}}\, T)/(I_{\text{ds}}\, E_{\text{EL}})] \times (I_{\text{R,EL}}/I_{\text{R,ref}})$$

Here, $I_{\text{R,EL}}$ is the spectrally integrated EL signal for a given ($V_{\text{ds}}$, $V_{\text{g}}$) bias condition, and $I_{\text{R,ref}}$ is the integrated detector signal from the QCL-illuminated Lambertian standard. All EQE values were calculated at each measured bias point using the corresponding $I_{\text{R,EL}}$ and $I_{\text{ds}}$ recorded from the $V_{\text{ds}}$-sweep (fixed $V_{\text{g}}$) and $V_{\text{g}}$-sweep (fixed $V_{\text{ds}}$) datasets. Any residual variation in the optical throughput or detector responsivity difference between $\lambda_{\text{ref}}$ and $\lambda_{\text{EL}}$ would only introduce a modest systematic error, so the extracted EQE values can be regarded as slightly conservative estimates.

## Supplementary Note 6. Spatially resolved EL mapping and 1064-nm reflection imaging

To identify the spatial distribution of mid-infrared EL emerging from the Te/MoS$_2$ heterojunction, we performed spatial mapping of the EL signal and compared it with a 1064-nm reflection map acquired over the same field of view. The sample was then raster-scanned by a motorized XY stage (step 1μm) while the collection optics were fixed. The output of the detector at each position was recorded after subtracting the background signal.

To enhance the spatial selectivity of the EL mapping, a circular aperture (0.85 mm diameter) was positioned at the center of the 1064-nm reflection spot in the collection path to define a spatially restricted detection region. Because the reflective objective is infinity-corrected, the beam between the objective and downstream optics is approximately collimated. In this configuration, the pinhole functions as an angular stop, defining an effective detection area on the sample. Under the paraxial approximation, its acceptance half-angle is $\theta_{\text{max}} \approx r/D$, where $r$ is the pinhole radius and $D$ is the objective-to-pinhole distance along the collimated beam. The corresponding effective detection diameter $d_{\text{eff}}$ on the sample can be estimated as:

$$d_{\text{eff}} \approx 2f_{\text{obj}}\theta_{\text{max}} \approx 2f_{\text{obj}}\frac{r}{D},$$

where $f_{obj}$ is the effective focal length of the objective. Using $r = 0.425$ mm, $D \approx 580$ mm and $f_{obj} \approx 5.0$ mm for the LMM40X-P01 objective, we obtain $d_{eff} \approx 7.3$ µm, which corresponds to the spatial resolution of the EL measurement.

We first acquired a 1064-nm reflection map to identify the device outline and metal contacts and to provide an unambiguous geometric reference under identical alignment. Without changing the optical alignment or scan coordinates, we subsequently recorded the mid-infrared EL map over the same area. For EL mapping, the emission was collected by the same reflective microscope objective (× 40), spectrally filtered by a 3.5-µm bandpass window (3500 ± 500 nm), and detected by an InSb detector. A reflective objective lens is used to avoid chromatic dispersion between the 1064 nm reflection wavelength and the 3.5 µm emission wavelength.

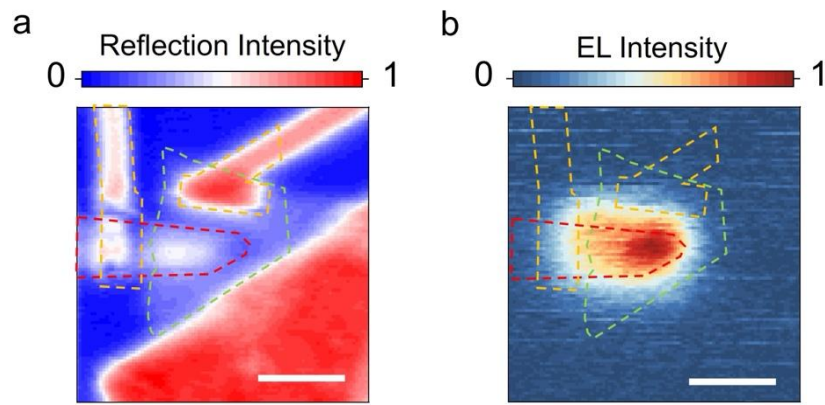

**Figure S5. Spatial mapping of mid-infrared EL and 1064-nm reflection maps.** **a,** 1064-nm reflection map acquired over the same field of view to identify the device outline and metal contacts. The green dashed outline marks $MoS_2$, the red dashed outline marks Te, and the yellow outline marks the metal electrodes. Scale bar: 30 µm. **b,** Integrated mid-infrared EL intensity map (3.0–4.0 µm bandpass) recorded under forward bias $(V_g = 20\,\text{V}, V_{ds} = 4\,\text{V}; 25\,\text{K})$. The EL map is compared with the 1064-nm reflection map in panel a. Scale bar: 30 µm.

**Supplementary Note 7. Measurement sensitivity in determining $V_{on}$**

In this Note, we clarify why the apparent EL turn-on voltage $V_{on}$ depends on the detection scheme. In the spectrally resolved measurement (Fig. 3b), the EL signal is collected through a monochromator, where the finite optical throughput and slit-limited signal-to-noise ratio reduce the detected photon flux and effectively raise the detection threshold. In contrast, in the high-sensitivity configuration (Fig. 3c), the integrated mid-infrared EL intensity is directly recorded by an InSb photodetector through a 3–4 µm bandpass filter, avoiding wavelength-selection losses and enabling earlier detection of weak emission. This integrated detection yields an apparent $V_{on} \approx 1.7$ V, lower than the onset inferred from spectrally resolved data. Similar sensitivity-dependent shifts of the apparent turn-on have been noted when EL is measured with more sensitive photodiode-based detection.

This discrepancy is expected because $V_{on}$, as defined in the main text, is the minimum bias at which the integrated EL signal clearly exceeds the background floor; it is therefore an operational metric that depends on measurement sensitivity and detection bandwidth. A lower apparent $V_{on}$ correspondingly reduces the electrical input required to reach detectable emission, which is beneficial for lowering power consumption at turn-on.

**Supplementary Note 8. Comparison of EL degree of polarization in representative Te/TMD heterojunctions**

Figure S6 compares the normalized EL and PL degree of linear polarization (DOP) of two representative Te/TMD heterojunction LEDs measured at 25 K, including a Te/MoS$_2$ device (red) and a Te/WSe$_2$ device (blue). Figure S6a shows the EL DOP, and Fig. S6b shows the PL DOP. While the drain bias and gate voltage conditions are not the same due to device variation on the turn-on behavior and the accessible operating windows are device dependent, we chose operating points with injection current densities that are as close as possible to enable a meaningful comparison. The DOP is normalized to the maximum value obtained for each device to highlight the relative polarization level.

For the Te/MoS$_2$ device, the DOP was measured at an injection current density of approximately 0.85 µA/µm$^2$ (using an effective emission area of about 300 µm$^2$ to estimate the current density). For the Te/WSe$_2$ device, the DOP was measured at an injection current density of approximately 0.88 µA/µm$^2$ with an effective area of ~440 µm$^2$. At these representative operating points, the corresponding absolute DOP values are ~0.70 for Te/MoS$_2$ and 0.98 for Te/WSe$_2$ (Fig. S6a).

To further examine whether the polarization variation is also reflected in optical excitation, we performed polarization-resolved micro-PL measurements on the corresponding Te flakes using an excitation power of 10 mW. The Te/WSe$_2$ device shows a nearly unity PL DOP of ~0.98. In contrast, the Te/MoS$_2$ device exhibits a much lower PL DOP of ~0.55 under our measurement conditions (Fig. S6b).

Consistent with the discussion in the main text, Fig. S6 illustrates that different Te flakes can exhibit distinct EL/PL polarization levels. A plausible interpretation is that differences in the intrinsic doping level among Te flakes may lead to variations in carrier distribution during electrical injection, which could contribute to the observed device-to-device variation in DOP.

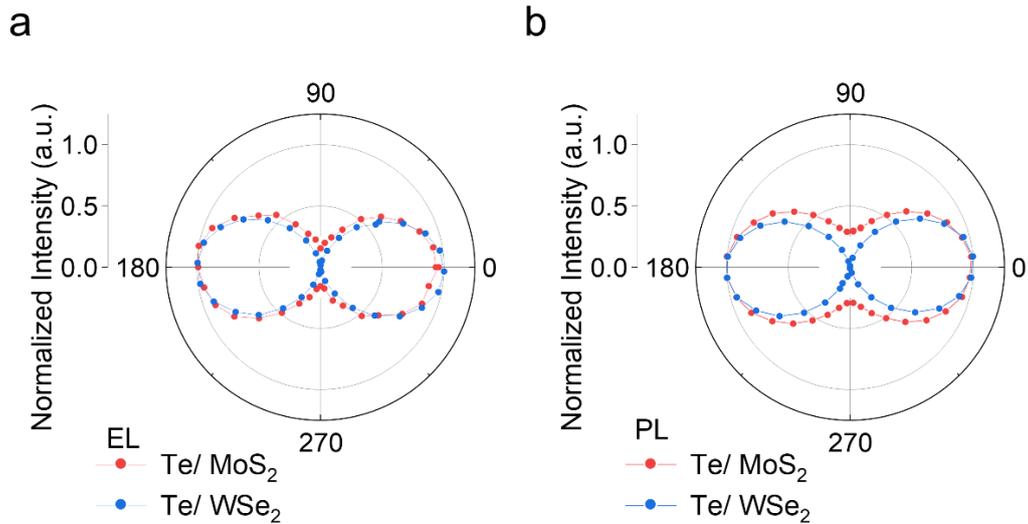

**Figure S6. Comparison of normalized degree of polarization (DOP) for two Te/TMD heterojunction emitters measured at 25 K.** Red symbols denote a Te/MoS$_2$ device and blue symbols denote a Te/WSe$_2$ device. The polarization values are normalized to each device's maximum DOP to emphasize comparative trends. (a) Normalized EL DOP measured at matched injection conditions, with current density of ~0.85 µA/µm$^2$ (~300 µm$^2$) for Te/MoS$_2$ and ~0.88 µA/µm$^2$ (~440 µm$^2$) for Te/WSe$_2$. (b) Normalized PL DOP measured on the corresponding Te/TMD heterojunction by polarization-resolved micro-PL at 25 K using an excitation power of 10 mW.

**Supplementary Note 9. Long-term stability of electroluminescence**

To assess the long-term stability of our Te/MoS$_2$ electroluminescent devices, we re-measured one representative device after prolonged storage and compared its EL characteristics with those obtained shortly after fabrication. The device was stored in an ambient condition for ~10 months and then re-characterized at 25 K using the same optical collection path and electrical measurement configuration. As shown in Figure S7, the integrated EL intensity as a function of $V_{ds}$ exhibits no obvious degradation after aging. The EL response and turn-on behavior remain essentially unchanged within experimental uncertainty. These results indicate that the Te/MoS$_2$ heterojunction maintains robust EL performance after long-term storage, supporting the practical stability of the present device platform.

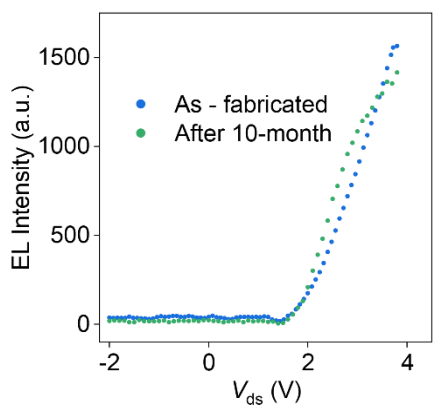

**Figure S7. $V_{ds}$-dependent integrated EL intensity measured at 25 K immediately after fabrication (blue) and 10 months later (green).** The device was stored in ambient conditions.

**Supplementary Note 10. Device-to-device reproducibility of gate-modulated electroluminescence**

In order to assess the reproducibility of gate-modulated mid-infrared electroluminescence in Te/TMD heterojunction LEDs, we fabricated and measured multiple devices based on similar Te/TMD architectures. Representative optical micrographs and gate-dependent EL responses of two independent devices are presented here to demonstrate that the observed emission behavior is not unique to a single sample, but is representative of the broader Te-based heterojunction system. In each optical image, the Te flake region is outlined by a white dashed line, the $MoS_2$ or $WSe_2$ flake is outlined by green dashed line, and the metal electrodes (Pd/Au or Cr/Au) are outlined by yellow dashed lines.

Figures S8a and S8b show measurements from a second Te/$MoS_2$ heterojunction LED. The optical micrograph in Fig. S8a shows the device geometry including the Te and $MoS_2$ flakes and the contact electrodes. Figure S8b plots the integrated mid-infrared EL intensity as a function of back-gate voltage measured at a fixed drain bias of 6 V under identical optical collection conditions. The EL intensity exhibits a clear rise and fall behavior with increasing $V_g$, qualitatively consistent with the gate-modulated EL response presented in the main text for the first Te/$MoS_2$ device. While the absolute intensity and the $V_g$ window for peak emission vary somewhat due to differences in flake geometry and contact configuration, the overall trend of gate-controlled EL remains consistent.

Figures S8c and S8d show measurements from an independent Te/$WSe_2$ heterojunction LED. The optical micrograph in Fig. S8c identifies the overlap region between the Te and $WSe_2$ flakes, as well as the contact electrodes. The corresponding integrated EL intensity versus $V_g$ data in Fig. S9d again display a gate-modulated response under the same drain bias of 6 V. The overall shape of the EL–$V_g$ curve is similar to those of the Te/MoS$_2$ devices, indicating that gate-controlled injection physics can be realized in Te/WSe$_2$ heterostructures as well.

These additional device results provide evidence that the gate-controlled injection physics and the associated electroluminescence characteristics can be reproduced in more than one device implementation. The inclusion of optical images with clear flake boundaries and EL response curves for multiple devices in the SI follows common practice in the emerging 2D emitter literature to demonstrate qualitative reproducibility across independently fabricated devices.

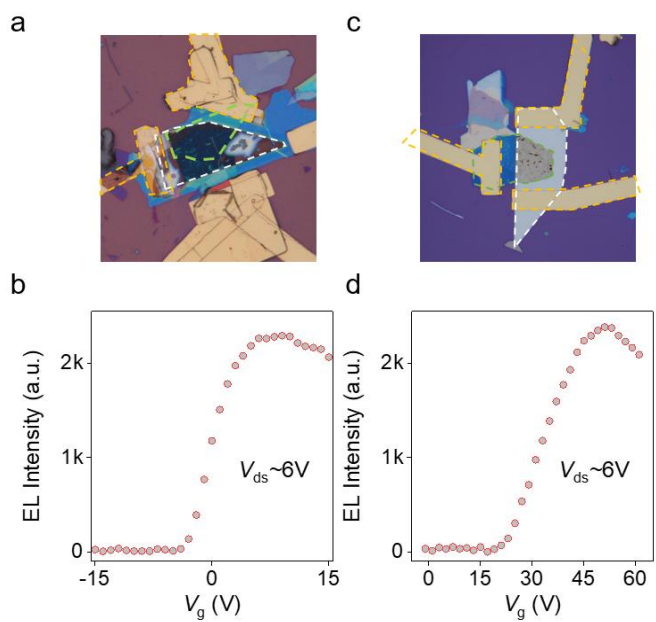

**Figure S8. Gate-voltage dependence of integrated EL intensity for multiple Te/TMD heterojunction LEDs measured at 25 K.** (a) Optical image of a Te/MoS$_2$ device; (b) integrated EL vs $V_g$ for the device in (a); (c) optical image of a Te/WSe$_2$ device; (d) integrated EL vs $V_g$ for the device in (c). Te flakes (white dashed), MoS$_2$/WSe$_2$ flakes (green dashed), and metal contacts (yellow dashed) are indicated.

**Supplementary Note 11. Gate-Dependent polarization stability in electroluminescence**

In addition to the intensity modulation of electroluminescence (EL) under back-gate control, we also examine the stability of the EL polarization state as a function of the back-gate voltage. Figure SX presents the degree of linear polarization (DOP) measured at 25 K for a representative Te/MoS$_2$ heterojunction LED under a fixed drain bias ($V_{ds}$ ~3.5 V). Panel (a) compares the normalized polarization patterns measured at two representative back-gate voltages ($V_g$ = 10 V and $V_g$ = 30 V).

Panel (b) summarizes the extracted DOP values as a function of the back-gate voltage from 5 V to 30 V, with error bars indicating the measurement uncertainty at each point.

The data show that, while the EL intensity can be strongly modulated by the back-gate voltage, the DOP remains essentially constant (~0.58–0.59) over the gate sweep. This result demonstrates that the EL emission in this Te/MoS$_2$ device exhibits polarization-locked intensity control: the polarization state is not significantly altered by the back-gate tuning that modulates injection efficiency. Together with the temperature-dependent EL data presented elsewhere in the Supplementary Information, these results support a coherent picture in which the EL intensity can be tuned by electrostatic control without altering the polarization state that is intrinsic to the Te flake and device configuration.

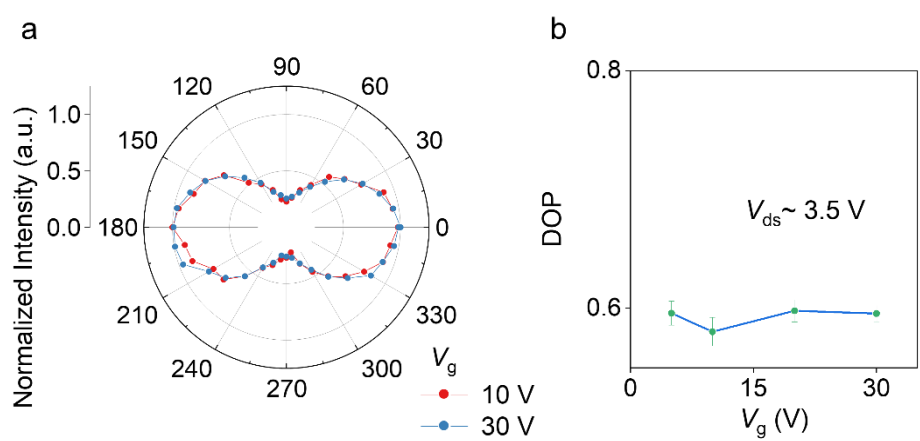

**Figure S9. Gate-voltage dependence of the degree of linear polarization (DOP) in a Te/MoS$_2$ heterojunction LED measured at 25 K with fixed drain bias ($V_{ds}$ ~3.5 V).** (a) Normalized polarization patterns measured at $V_g$ = 10 V (red) and $V_g$ = 30 V (blue); (b) Extracted DOP as a function of back-gate voltage from 5 V to 30 V, with error bars indicating measurement uncertainty.

**Supplementary Note 12. Linewidth analysis of bias- and gate-dependent electroluminescence**

We extracted the full width at half maximum from all EL spectra using the same half-maximum method. The error bars in Fig. S10 represent the combined uncertainty from statistical errors and noise-induced errors. As shown in Fig. S10a, for the bias-dependent EL spectra measured at fixed $V_g$ = 20 V, the full width at half maximum decreases from about 17 meV at $V_{ds}$ = 3.0 V to about 12 meV at $V_{ds}$ = 4.0 V, indicating a weak linewidth narrowing trend with increasing forward bias. However, the magnitude of this change remains relatively small compared with the absolute linewidth itself. This narrowing trend does not mainly originate from a pronounced shift of the main emission peak. Instead, it is primarily associated with the reduced relative contribution from the low-energy side of the asymmetric EL spectrum. To quantify this asymmetry, we further decompose the linewidth as FWHM

= FWHM$_l$ + FWHM$_h$, where FWHM$_l$ (FWHM$_h$) denotes the energy width from the peak position to the half-maximum point on the low-energy (high-energy) side. As shown in Fig. S10a, FWHM$_l$ decreases noticeably with increasing $V_{ds}$ (or emission intensity), whereas FWHM$_h$ remains relatively unchanged. At low bias and under weak emission conditions, the low-energy-side intensity makes a noticeable contribution, which shifts the left half-maximum point toward lower energy and results in a larger apparent full width at half maximum. With increasing $V_{ds}$, the emission becomes increasingly dominated by band-edge recombination, while the relative contribution from the low-energy-side emission decreases. The low-energy in-gap emission may originate from shallow impurity states, the Franz-Keldysh effect induced by the built-in electric field, and excitonic emission. Under stronger excitation, these contributions gradually become saturated or are masked by the stronger band-edge emission.To evaluate the effect of gate control on the EL linewidth, we performed the same linewidth analysis on the gate-dependent EL spectra shown in Fig. 4d. As shown in Fig. S10b, for the gate-dependent EL spectra measured at a fixed forward-bias condition of ~4 V, the full width at half maximum first decreases from about 13 meV at $V_g$ = 5 V to about 10 meV at $V_g$ = 20 V, where the EL emission is strongest, and then slightly increases to about 12 meV at $V_g$ = 30 V. This overall variation remains small compared with the absolute linewidth itself. Notably, a narrower linewidth is correlated with stronger EL emission. This trend is consistent with the bias-dependent behavior in Fig. S10a. Therefore, the weak linewidth evolution with gate voltage is more naturally attributed to changes in the relative contribution of the low-energy-side emission under different injection conditions.

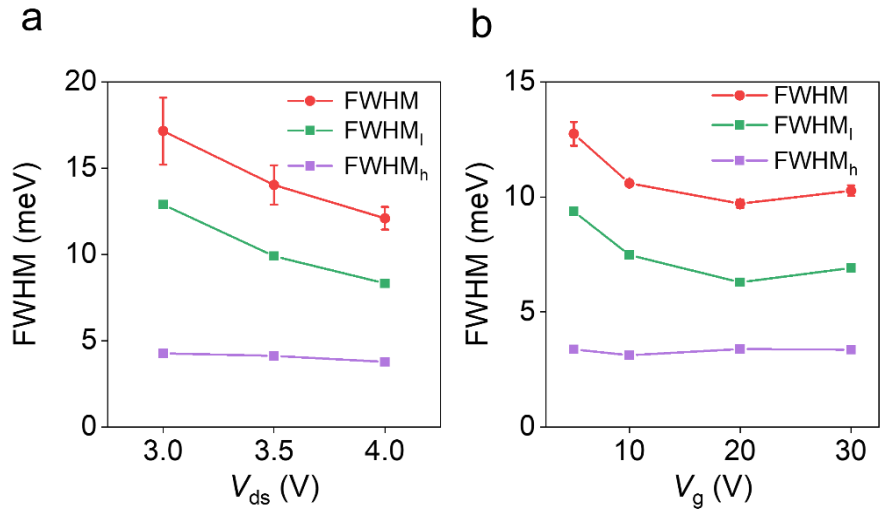

**Figure S10. Extracted EL full width at half maximum under bias and gate modulation.** (a) FWHM of the bias-dependent EL spectra measured at $V_{ds}$ = 3.0, 3.5, and 4.0 V, with $V_g$ fixed at 20 V. (b) FWHM of the EL spectra measured under different back-gate voltages at a fixed forward-bias ~4 V. FWHM (red), $FWHM_l$ (green), and $FWHM_h$ (purple) are plotted, where $FWHM_l$ and $FWHM_h$ denote the half widths from the peak energy to the half-maximum points on the low-energy and high-energy sides, respectively. The error bars represent the combined uncertainty from statistical errors and noise-induced errors.